\documentclass[twocolumn,prb,showpacs,preprintnumbers,amsmath,amssymb]{revtex4}
\usepackage{graphicx}% Include figure files
\usepackage{dcolumn}% Align table columns on decimal point
\usepackage{bm}% bold math

\begin{document}

%\preprint{APS/123-QED}

\title{Zn-impurity effects on quasi-particle scattering in La$_{2-x}$Sr$_x$CuO$_4$ studied by angle-resolved photoemission spectroscopy}

\author{T. Yoshida$^1$, Seiki Komiya$^2$, X. J. Zhou$^3$, K. Tanaka$^3$, A.
Fujimori$^1$, Z. Hussain$^4$, Z.-X. Shen$^3$, Yoichi Ando$^5$, H.
Eisaki$^6$, S. Uchida$^1$} \affiliation{$^1$Department of Physics,
University of Tokyo, Bunkyo-ku, Tokyo 113-0033, Japan}
\affiliation{$^2$Central Research Institute of Electric Power
Industry, Komae, Tokyo 201-8511, Japan}\affiliation{$^3$Department
of Applied Physics and Stanford Synchrotron Radiation Laboratory,
Stanford University, Stanford, CA94305} \affiliation{$^4$Advanced
Light Source, Lawrence Berkeley National Lab, Berkeley, CA 94720}
\affiliation{$^5$Institute of Scientific and Industrial Research,
Osaka University, Ibaraki, Osaka 567-0047, Japan}
\affiliation{$^6$National Institute of Advanced Industrial Science
and Technology, Tsukuba 305-8568, Japan}
\date{\today}% It is always \today, today,

\begin{abstract}
Angle-resolved photoemission measurements were performed on
Zn-doped La$_{2-x}$Sr$_x$CuO$_4$ (LSCO) to investigate the effects
of Zn impurities on the low energy electronic structure. The
Zn-impurity-induced increase in the quasi-particle (QP) width in
momentum distribution curves (MDC) is approximately isotropic on
the entire Fermi surface and energy-independent near the Fermi
level ($E_F$). The increase in the MDC width is consistent with
the increase in the residual resistivity due to the Zn impurities
if we assume the carrier number to be 1-$x$ for $x$=0.17 and the
Zn impurity to be a potential scatterer close to the unitarity
limit. For $x$=0.03, the residual resistivity is found to be
higher than that expected from the MDC width, and the effects of
antifferomagnetic fluctuations induced around the Zn impurities
are discussed. The leading edges of the spectra near ($\pi$,0) for
$x$=0.17 are shifted toward higher energies relative to $E_F$ with
Zn substitution, indicating a reduction of the superconducting
gap.
\end{abstract}

%\keywords{Suggested keywords}%Use show keys class option if keyword
\maketitle
\section{Introduction}
%background of the study of impurity effects
Zn substitution for Cu atoms in the CuO$_2$ planes for the
high-$T_c$ cuprates causes a dramatic reduction of $T_c$ and thus
may offer an opportunity to characterize the nature of the
superconducting states in the cuprates. Zn is a nonmagnetic
impurity with a closed $d$ shell and produces a large in-plane
residual resistivity while the temperature-slope of the
resistivity is unchanged.\cite{Fukuzumi} Quantitative analysis of
the residual resistivity indicates that the Zn impurity acts as a
potential scatterer in the unitarity limit. To understand the
microscopic mechanism of the scattering by the Zn impurities, the
local information around the Zn impurities has been extensively
studied. Reduction of superfluid density by Zn impurities were
detected by $\mu$SR \cite{Nachumi} and a ``Swiss cheese" model, in
which non-superconductivity islands are induced around the Zn
impurities, was proposed to explain the reduction of the
superconducting fraction proportional to the Zn concentration.
This model is consistent with the observation by scanning
tunneling microscopy (STM) that the superconductivity is locally
destroyed by Zn impurities.\cite{PanSTM} Also, according to NMR
studies, it was found that antiferromagnetic moment is induced
around the Zn impurity.\cite{Julien} While these experimental
results have given insight into the local electronic structure
around the impurity, Zn impurity effects in momentum space, which
are more directly related to the transport properties, is still
poorly understood.

%previous ARPES results of impurity effects
In order to elucidate details of impurity scattering in momentum
space as well as the mechanism of the reduction of $T_c$ induced
by impurity, direct observation of the quasi-particle (QP) under
the influence of Zn impurities should give useful information.
Previous angle-resolved photoemission (ARPES) studies indicated
that in Bi$_2$Sr$_2$CaCu$_2$O$_{8+\delta}$(Bi2212) Zn and Ni
impurities reduces the coupling strength to Boson excitation which
leads the ``kink" in the QP dispersion.\cite{Terashima,
Zabolotnyy} Also, the depression of the superconducting coherence
peak by Zn impurities has been observed,\cite{Terashima,
Zabolotnyy, TerashimaZnLSCO, Nishina} consistent with the decrease
in the superfluid density.\cite{Nachumi} In this work, we shall
focus on the impurity effects on the QP in La$_{2-x}$Sr$_x$CuO$_4$
(LSCO), particularly, on the relationship between the $T_c$ and
the observed increase in the widths of momentum distribution
curves (MDC) in relation to the scattering mechanism of QP in the
normal state. In particular, the effects of antiferromagnetic
fluctuations locally induced by the Zn impurities shall be
discussed.

\begin{figure}
\includegraphics[width=9cm]{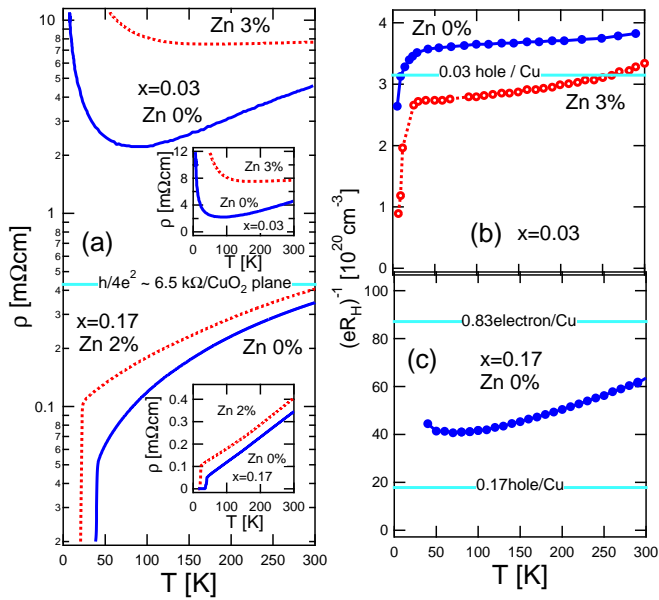}
\caption{\label{sample}(Color online) Transport properties of
Zn-doped LSCO samples studied in this work. (a): Electrical
resistivity $\rho$. Insets show $\rho$ on linear scale. (b),(c):
Hall coefficient $R_H$. }
\end{figure}

\section{Experiment}
%Experiment
High-quality single crystals of Zn-doped and Zn-free LSCO were
grown by the traveling-solvent floating-zone method. Figure
\ref{sample} shows the temperature dependence of the electrical
resistivity $\rho$ and the Hall coefficient $R_H$ of samples
studied in the present work. The Zn-free and 2 \% Zn-doped samples
with hole content of $x$ = 0.17 have critical temperatures
($T_c$'s) of 40 K and 22 K, respectively. The Zn-free and 3 \%
Zn-doped $x$ = 0.03 samples are non-superconducting. With
Zn-doping, the residual resistivity increases for both $x$=0.03
and 0.17 samples, while the temperature slope of $\rho$ does not
change with the Zn impurities for the $x$=0.17 sample. The
``carrier number" defined by $(eR_H)^{-1}$ for $x$=0.03 is close
to the nominal hole concentration, but decreases by $\sim$ 20 \%
with 3\% Zn-doping. The ARPES measurements were carried out at
beamline 10.0.01 of Advanced Light Source (ALS) and beamline 5-4
of Stanford Synchrotron Radiation Laboratory (SSRL), using
incident photons with energies of 55.5 eV and 22.4 eV,
respectively. The total energy resolution was about 15 meV (SSRL)
or 20 meV (ALS). The momentum resolutions at ALS and SSRL are
0.02$\pi$ and 0.01$\pi$ in units of 1/$a$, respectively, where
$a$=3.8 \textrm{\AA} is the lattice constant. The samples were
cleaved \textit{in situ} and measurements were performed at about
20 K (ALS) or 10 K (SSRL). In the measurements at ALS, the
electric vector $\mathbf{E}$ of the incident photons lied within
the CuO$_2$ plane, 45 degrees rotated from the Cu-O direction and
was parallel to the Fermi surface segment in the nodal region.
This measurement geometry enhances dipole matrix elements in this
$ \mathbf{k}$ region because the wave function has $x^2-y^2$
symmetry.\cite{yoshidaOD} The $\mathbf{E}$ in the measurements at
SSRL was nearly parallel to the Cu-O direction.

\begin{figure}
\includegraphics[width=8cm]{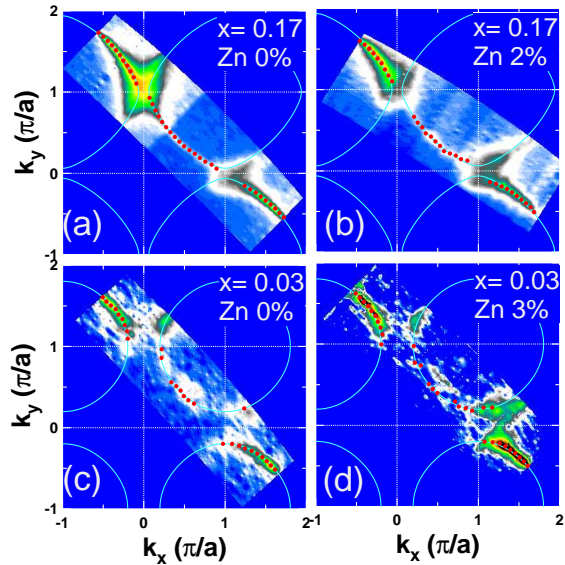}
\caption{\label{FS}(Color online) $k$-space spectral weight
mapping at $E_F$ for La$_{2-x}$Sr$_x$Cu$_{1-y}$Zn$_y$O$_4$. Dots
indicate $k_F$ positions determined by MDC peaks at $E_F$. Solid
curves show the Fermi surface fitted to the tight-binding model.}
\end{figure}

\section{Results and discussion}
\subsection{Zn-impurity effects on the quasi-particle}
%Fermi surface
First, in order to see the effects of Zn impurities on the shape
of the Fermi surface, momentum-space spectral weight mapping at
the Fermi level ($E_F$) is shown in Fig. \ref{FS}. We have
determined the Fermi momentum ($k_F$) of the (underlying) Fermi
surfaces by using the peak positions of the MDC's, as represented
by dots.\cite{shadow} The $k_F$ positions thus determined, which
designate the shape of the Fermi surfaces, could be well fitted to
the Fermi surface of the tight-binding model as shown by solid
curves. Both for $x$=0.03 and 0.17, the tight-binding parameters
are almost identical between the Zn-doped and Zn-free samples
($t^\prime/t$=-0.120, $t^{\prime\prime}/t=0.10$ for $x$=0.03 and
$t^\prime/t$=-0.135, $t^{\prime\prime}/t=0.075$ for $x$=0.17),
indicating that the Zn substitution does not change the overall
electronic structure and the hole concentration appreciably.
Although $R_H$ for $x$=0.03 shown in Fig. \ref{sample}(b) suggests
a decrease of the hole concentration with Zn doping by $\sim$20\%,
we could not detect this signature in the present experiment since
the corresponding change  is a very small portion of the Fermi
surface area $1+x$ ($\ll$1\%).

\begin{figure*}
\includegraphics[width=15cm]{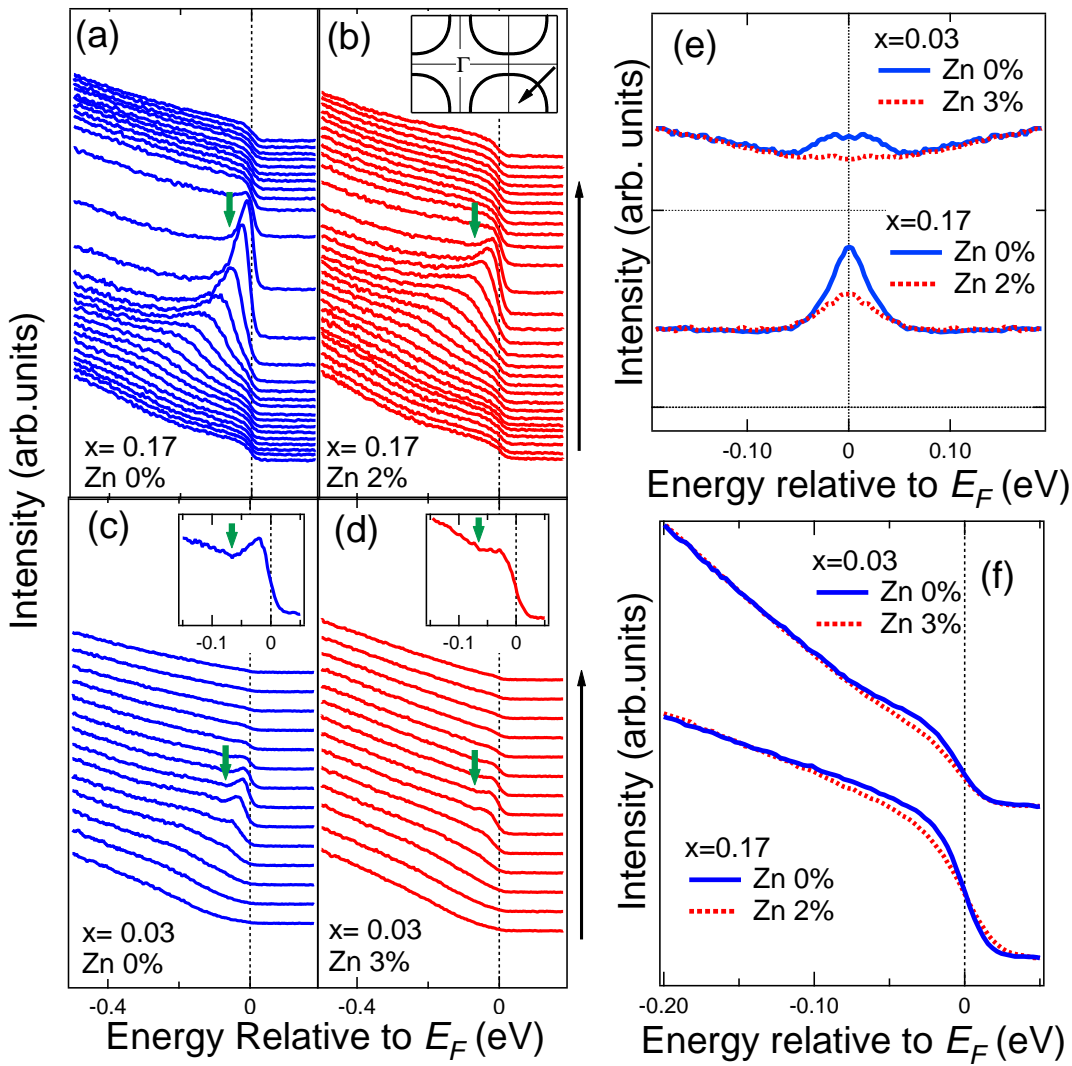}
\caption{\label{EDC}(Color online) ARPES spectra of
La$_{2-x}$Sr$_x$Cu$_{1-y}$Zn$_y$O$_4$ in the nodal direction of
the second Brillouin zone. (a)-(d): EDC's  corresponding to the
cut shown by an arrow in the inset. Arrows in the panels indicate
the energy positions of a dip in the spectral line shape. Insets
in (c) and (d) show EDCs near $k_F$ with clear dips. (e):
Symmetrized EDC's at $k_F$ with respect to the $E_F$. (f): EDC's
integrated along the nodal direction. The spectra in (e) and (f)
have been normalized to the integrated spectral weight between
-0.2- -0.1 eV.}
\end{figure*}

%EDC's in node direction
Nevertheless, Zn-impurity effects are clearly seen in the spectral
line shapes of the energy distribution curves (EDC's). Figure
\ref{EDC} shows EDC's in the nodal direction [(0,0)-($\pi$,$\pi$)]
in the second Brillouin zone. Clear QP peaks are observed near
$E_F$ in the Zn 0\% $x$=0.17 and 0.03 samples [panels (a) and (c),
respectively]. With Zn substitution, the peak is significantly
depressed [panels (b) and (d)]. As indicated by vertical arrows, a
characteristic dip at $\sim$ 70 meV, which corresponds to the
``kink" is prominent for the $x$=0.03 spectra. The dip feature
still remains at almost the same binding energy with Zn-doping.

%Spectral weight
To clarify the impurity effects on the QP spectral weight, the
EDC's at $k_F$ have been symmetrized with respect to $E_F$ as
shown in Fig. \ref{EDC}(e). These spectra have been normalized to
the spectral weight between -0.2- -0.1 eV below the $E_F$.  For
both the Zn-free and Zn-doped $x$=0.17 samples, the symmetrized
EDC's show a clear QP peak at the $k_F$ point although the QP
intensity decreases with Zn-doping. The existence of a clear QP
peak even for the Zn-doped sample may be related with the
persistence of superconductivity with $T_c$ = 22 K. On the other
hand, the QP peak is strongly depressed in the Zn-doped $x$=0.03
sample. Particularly, the symmetrized EDC at $k_F$ show a shallow
dip at $E_F$, indicating the destruction of the nodal QP and a
possible pseudogap opening. Note that the pseudogap on the energy
scale of $\sim$10 meV caused by charge localization was observed
in the optical conductivity measurement of LSCO ($x$=0.03) at low
temperature.\cite{Dumm} The observed pseudogap-like feature of
similar magnitude in the EDC of the $x$=0.03 Zn 3 \% sample would
therefore be related to the localization behavior at low
temperature of the electrical resistivity as seen in Fig.
\ref{sample}. The integrated MDC spectra along the nodal direction
in Fig. \ref{EDC}(f) for both hole concentrations also show
suppression of the spectral weight by the Zn impurities.

\begin{figure}
\includegraphics[width=9cm]{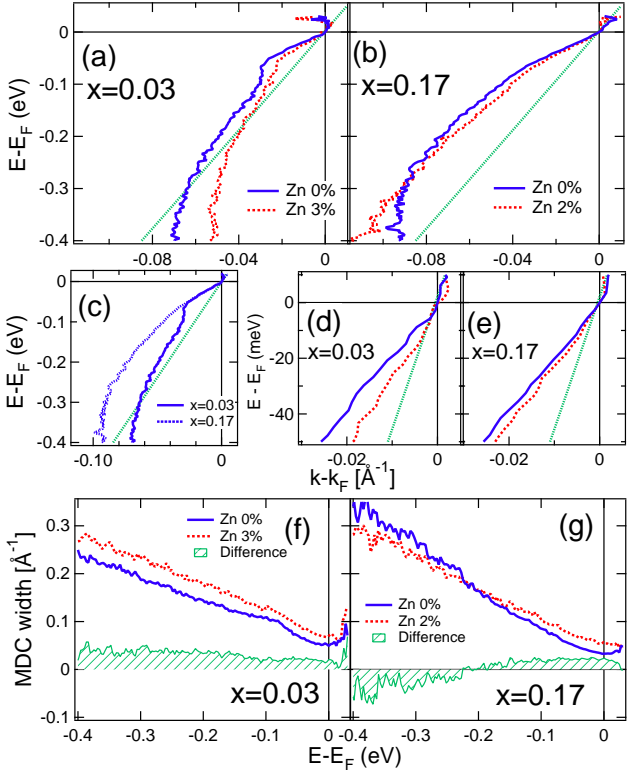}
\caption{\label{disp} (Color online) Quasi-particle peak in MDC's
in the nodal direction of La$_{2-x}$Sr$_x$Cu$_{1-y}$Zn$_y$O$_4$.
(a),(b): Energy dispersions determined by MDC peaks. (c):
Comparison between the Zn-free $x$=0.03 and 0.17 samples. (d),(e):
Enlarged plot near the $E_F$ of panels (a) and (b), respectively.
(f),(g): Energy dependence of the MDC width in the nodal
direction.}
\end{figure}

%Energy dispersions and MDC width in the node direction
In Fig. \ref{disp}, we summarize the dispersion of the QP peak in
MDC's and the energy dependence of the MDC width in the nodal
direction. Both the $x$= 0.03 and 0.17 samples show a kink at
$\sim$ 70 meV [panels (a),(b)]. In Fig. \ref{disp}(c), one can
confirm that the Fermi velocity for the Zn-free samples does not
change with hole doping, i.e., the universal Fermi
velocity.\cite{ZhouNature} By contrast, the slopes of the energy
dispersions near $E_F$ slightly increases with Zn-doping as shown
in panels (a) and (b). Panel (d) indicates that the slope of the
dispersion within $\sim$ 20 meV of $E_F$ becomes steeper  in $x$=
0.03 Zn 3\%, which however may be an artifact due to the (pseudo)
gap opening [Fig.\ref{EDC}(d)]. Indeed, such an effect is absent
in the $x$=0.17 samples [panel(e)].

\begin{figure}
\includegraphics[width=9cm]{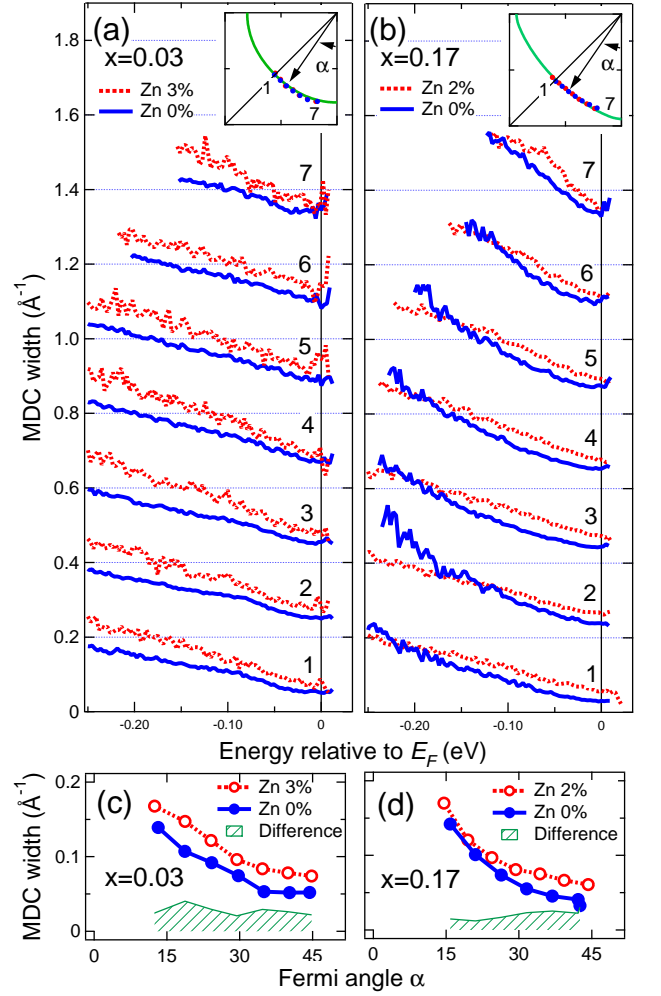}
\caption{\label{MDCAll}(Color online) MDC widths at various $k_F$
points on the Fermi surface of
La$_{2-x}$Sr$_x$Cu$_{1-y}$Zn$_y$O$_4$. (a),(b): Energy dependence
of the MDC width for $x$=0.03 and 0.17, respectively. The data are
shifted by 0.2 \AA$^{-1}$ between different momenta. (c),(d): MDC
width at $E_F$ as a function of Fermi angle $\alpha$. Differences
in the width between the Zn-free and Zn-doped samples are also
plotted.}
\end{figure}

%MDC width
Next, let us look at the effect of Zn impurities on the MDC width
in the nodal direction [Figs. \ref{disp}(f) and \ref{disp}(g)].
The MDC width shows a characteristic drop below the kink energy of
$\sim$70 meV, especially for the $x$=0.03 samples, indicating a
reduction of the scattering rate of QP by collective modes below
the mode energy. Below the kink energy, the Zn-induced increase in
the MDC width is energy independent as clearly indicated by the
difference between the Zn-doped and Zn-free data. Since the MDC
width $\Delta k$ increases with decreasing mean-free path $l$
through $\Delta k=1/l$ the extra scattering due to the impurity,
the energy-independent increase in the MDC width is consistent
with the temperature-independent increase in the resistivity
caused by the Zn impurities.

%MDCs along Fermi surface
In order to see the momentum dependence of the increase in the MDC
on the Fermi surface, the MDC widths at various $k_F$'s are shown
in Fig. \ref{MDCAll}. The energy dependence of the MDC width is
compared between the Zn-doped and Zn-free samples in Figs.
\ref{MDCAll}(a) and \ref{MDCAll}(b), and the MDC width at $E_F$ is
plotted in Figs. \ref{MDCAll}(c) and \ref{MDCAll}(d) as a function
of the Fermi angle $\alpha$. Also, the differences of the MDC
width between the Zn-free and Zn-doped samples are plotted on the
same panels. These plots indicate that the increase in the MDC
width by the Zn impurity is almost isotropic on the Fermi surface.
Thus, on the low energy scale ($<$ 70 meV), the contribution of
the impurity to the MDC width is energy- and momentum-independent,
indicating that the Zn impurity is a nearly static and isotropic
scatterer. As a results of the momentum independent increase in
the MDC width, the relative increase in the MDC width for 2-3\% Zn
substitution is 10-20\% in the anti-node direction and 50-100\% in
the node direction. Because the pseudogap size mainly determines
the $c$-axis transport properties in underdoped samples, the small
relative increase in the MDC width in the antinodal region
compared to the nodal region may explain the fact that the
out-of-plane transport is less affected by the Zn impurities than
the in-plane transport.\cite{Mizuhashi}

\subsection{Comparison with transport properties}
%Drude formula
Since the MDC width is equal to the inverse of the mean free path,
here, we shall compare the transport properties with the MDC width
deduced from the present ARPES results. The mobility
$\mu=e\tau/m^*$ can be derived from the in-plane resistivity
$\rho=m^*/ne^2\tau$ and the carrier density $n$ if the Drude
formula is assumed. We have assumed that $n$ is given by $n=x$ and
$n=1-x$ for $x$=0.03 and 0.17, respectively, because the Hall
coefficient\cite{AndoHall,OnoHall} $R_H=1/ne$  as well as the
residual resistivity of Zn-doped LSCO\cite{Fukuzumi} show a
crossover from $n \sim x$ to $n \sim 1-x$ around $x$=0.1. From the
ARPES data, the inverse mobility can be calculated using the
formula $\mu^{-1}=m^*/e\tau=m^*v_F/ev_F\tau=\hbar k_F \Delta k
/e$, where $\Delta k$ is the MDC width. As the observed MDC width
is influenced by the finite angular and energy resolutions, we
have subtracted the angular broadening $\sim$ 0.01 \AA$^{-1}$ from
the measured MDC widths. When the Fermi surface has a pseudo-gap
as in the underdoped region, since the QP's in the node region
dominate the in-plane transport, the evaluated $\mu^{-1}$ would be
that of the node region. As for the slightly overdoped $x$=0.17
samples, too, the scattering rate is the lowest in the nodal
region, as can be seen from the smallest MDC width in this region
[see Fig. \ref{MDCAll}(d)]. Therefore, the evaluated $\mu^{-1}$ is
largely determined by the nodal QP's.

\begin{figure}
\includegraphics[width=9cm]{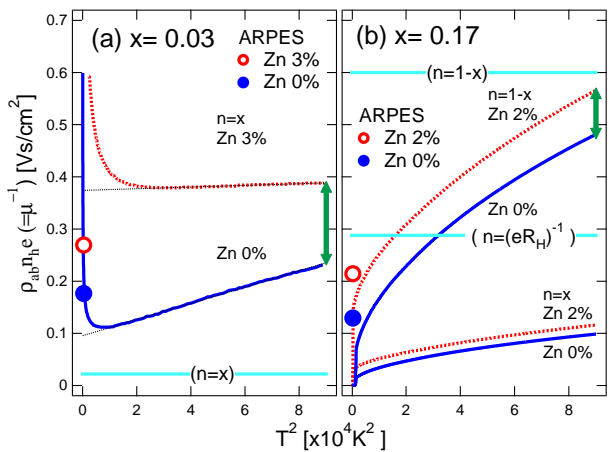}
\caption{\label{transport}(Color online) Comparison of the inverse
mobility obtained from the electrical resistivity and those
estimated from ARPES results for
La$_{2-x}$Sr$_x$Cu$_{1-y}$Zn$_y$O$_4$. Inverse mobility
corresponding to the universal resistance $h/4e^2$ and assumed
carrier number ($n$ in parentheses) are shown by horizontal lines.
Arrows indicate the increase in the inverse-mobility at $T$=300K,
which are compared with the ARPES results.}
\end{figure}

%Comparison with the transport results
In Fig. \ref{transport}, we compare the Zn-induced increase in the
inverse mobility $\mu^{-1}$ obtained from the transport and
present ARPES results.\cite{mdc} From the measurements of the
electrical resistivity $\rho$ of the present samples, the
inverse-mobility $\mu^{-1}=ne\rho$, with $n=x$ ($x$=0.03) or
$n=1-x$ ($x$=0.17), is found to increase with Zn doping by 0.16
and 0.098 Vs/cm$^2$ for $x$=0.03 and 0.17, respectively, at 300 K.
From the MDC width at $E_F$ in the nodal direction (Fig.
\ref{disp}), the Zn-induced increase in $\Delta k$ are 0.022 and
0.020 \AA$^{-1}$ for the $x$= 0.03 and 0.17 samples, respectively.
Using the increase in $\Delta k$ and the formula $\mu^{-1}=\hbar
k_F \Delta k /e$, the increase in $\mu^{-1}$ for $x$=0.03 and 0.17
samples are calculated to be 0.093 and 0.085, respectively. Here,
$k_F\sim$ 0.64 \AA$^{-1}$ measured from ($\pi,\pi$) was deduced
from ARPES. Therefore, for the $x$=0.17 samples, the obtained
Zn-induced increase in the $\mu^{-1}$ values from the transport
and ARPES results are quantitatively consistent if we assume
$n=1-x$. On the other hand, for the $x$=0.03 samples, the increase
in $\mu^{-1}$ from ARPES is smaller than that estimated from the
transport.

%Difference between transport and ARPES
In general, transport and ARPES measurements detect different
scattering processes of QP in different weight. The scattering
rate $1/\tau_{\mathrm{tr}}$ by impurities in transport is given by
$1/\tau_{\mathrm{tr}}\propto \frac{1}{2}\int
v(\mathbf{k}-\mathbf{k^\prime})(1-\cos\theta)d\theta$, where
$\theta$ is the scattering angle and $v(\mathbf{q})$ is the
scattering potential.\cite{Mahan} Because of the factor
1-$\cos\theta$ in the integrand, the main scattering process stems
from backward scattering. On the other hand, the scattering rate
in ARPES ($1/\tau_{\mathrm{ARPES}}$) is
$1/\tau_{\mathrm{ARPES}}=2Z{\rm Im}\Sigma\propto \frac{1}{2}\int
v(\mathbf{q})d\theta$, where $\Sigma$ is the self-energy and $Z$
is the renormalization factor, indicating that both backward
scattering and forward scattering equally contribute. It should be
noted that, in the unitarity limit, $1/\tau_{\mathrm{tr}}$ and
$1/\tau_{\mathrm{ARPES}}$ should be the same due to the
$\mathbf{q}$-independence of the scattering amplitude. Since the
Zn-induced increase in $\mu^{-1}$ from the transport (assuming
$n=1-x$) and ARPES results for $x$=0.17 samples show almost the
same value of $\sim0.09$ Vs/cm$^2$, one can conclude that the
scattering by the Zn impurities in the $x$=0.17 samples are in the
unitarity limit. This is consistent with the in-plane resitivity
results\cite{Fukuzumi} and with the conclusion reached by an STM
study.\cite{PanSTM}

%interpretation of the comparison between transport and ARPES
For the $x$=0.03 samples, on the other hand, the increase in
$\mu^{-1}$ from ARPES is smaller than that from transport (which
is extrapolated from the high temperature region in order to
eluminate the localization effect). This cannot be understood
within the unitary scattering picture. Here, let us consider the
effects of antiferromagnetic correlations induced around the
impurity atom.\cite{Julien} Theoretically, vertex correction from
the antiferromagnetic fluctuations may cause an enhancement of the
resistivity compared to that calculated based on Boltzmann
equation.\cite{Kontani} Since the present estimate from the ARPES
data is based on Boltzmann transport theory, the smaller
$\mu^{-1}$ estimated from ARPES than that from the transport
indicates that the QP scattering rate has a peculiar momentum
dependence with a maximum occuring in the backward scattering
direction. Such a situation may be realized if scattering by the
low-energy antiferromagnetic fluctuations induced by the Zn
impurities is dominant in the QP scattering in the $x$=0.03
samples.

%Mechanism of the reduction of superconductivity
In two dimensional metals like high-$T_c$ cuprates, whether a
dirty system become a superconductor or not is determined by the
normal-state critical sheet resistance.\cite{Emery} In both Zn 0\%
and 3\% $x$=0.03 samples, it is clear that the inverse mobilities
obtained by the ARPES and transport studies are much higher than
critical universal 2D value, and therefore, the system indeed goes
into the localization phase. This explains the absence of
superconductivity in the Zn-free $x$=0.03 samples, although it
shows a clear QP in the vicinity of $E_F$ near the node direction.
It is likely that a tiny gap is opened in the nodal direction
[Fig. \ref{EDC}(e)], as pointed out in previous report on the
lightly-doped cuprates.\cite{KMShenPRB} With Zn-doping, the QP
peak is significantly depressed as shown in Figs. \ref{EDC}(e) and
\ref{EDC}(f), a signature of carrier localization which starts at
higher temperature. Note that a pseudogap on the low energy scale
of $\sim$10 meV was observed in the optical conductivity
measurement and indicated localization at low
temperatures.\cite{Dumm} In the Zn 2\% $x$=0.17 samples, the
inverse mobility obtained by ARPES are lower than the universal
resistivity, which accords with the persistence of the
superconductivity in this sample. One can infer that, with a few
more percent Zn-doping, the inverse mobility would reach that of
the universal resistivity and the system would enter the
non-superconducting phase.

\begin{figure}
\includegraphics[width=9cm]{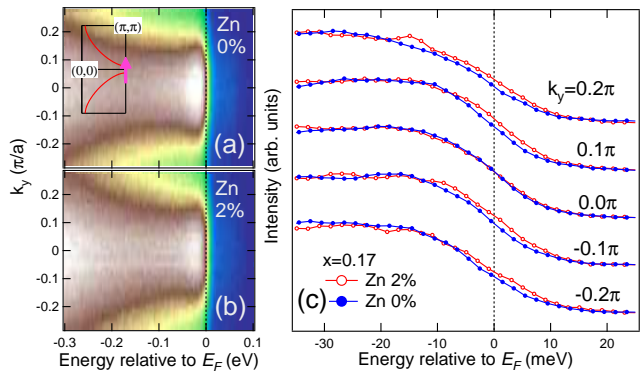}
\caption{\label{pi0EDC}(Color online) Zn impurity effects on the
spectra around ($\pi$,0) for LSCO $x$=0.17 samples. (a) and (b)
show intensity plots in $E-k_y$ space (with $k_x$=$\pi$ fixed) for
Zn 0\% and 2\%, respectively. Inset shows cut direction. (c):
Comparison of the EDC's of Zn 0\% and 2\% samples. The EDC's at
$k_y=\pm0.1\pi$ show energy-shifts of the spectra with Zn doping.}
\end{figure}

\subsection{Zn-impurity effects in the anti-nodal region}
%(pi,0) energy gap
Finally, we discuss Zn impurity effects on the spectra in the
antinodal region, where the opening of the superconducting gap is
expected below $T_c$. Figure \ref{pi0EDC} shows EDC's along the
($\pi$,0)-($\pi$,$\pi$) line of the $x$= 0.17 samples measured at
SSRL. At the $k_F$ point with $k_y\sim \pm 0.1\pi$, the EDCs of
the Zn-doped sample are slightly shifted toward higher energies
relative to $E_F$ compared to the Zn-free sample. Because the
pseudogap state has been reported to be stabilized by out-of-plane
disorder,\cite{Okada} if in-plane and out-of-plane disorder have
similar effects on the pseudogap, the observed shift is unlikely
to be a reduction of the pseudogap. Instead, the shift may be
understood as a reduction of the superconducting gap, because such
a reduction by Zn impurity has also been observed in Zn-doped
Bi2212.\cite{Nishina} However, the decrease of the gap by
Zn-doping is only 1-2 meV, which is much smaller than that ($\sim$
4 meV) expected from the decrease of $T_c$ by 20 K. Furthermore,
the observed gap size is too small compared with that estimated
from the $d$-wave BCS theory $\Delta\sim$ 8 meV for the Zn-free
samples with the $T_c \sim$ 40 K.\cite{Maki} Note that, for
$x$=0.15 samples, an indication of the leading edge gap of $\sim$
10 meV was reported in previous studies although no clear
coherence peak was seen around $(\pi,0)$\cite{TerashimaLSCO} even
using higher energy resolution than the present one. Therefore,
the superconducting gap identified in this study may be opened on
top of the pseudo gap and hence is probably blurred compared to an
ideal superconducting gap. More detailed systematic studies are
necessary to characterize the weak superconductivity in the
antinodal region.

\section{Conclusion}
%Conclusion
In summary, we have studied Zn-impurity effects on the near $E_F$
electronic states of LSCO and discussed their relationship to the
transport properties. We have observed an isotropic increase in
the MDC width as well as the suppression of spectral weight in the
low energy part of the spectra. For slightly overdoped $x$=0.17,
the increase in the MDC width is close to that expected from the
unitary limit of the impurity scattering and explains the increase
in the in-plane residual resistivity. For the lightly doped
$x$=0.03, we found that the residual resistivity is larger than
that expected from the MDC width. We propose that backward
scattering  due to antiferromagnetic fluctuations may be enhanced
compared to forward scattering. We have confirmed that
superconductor-to-localization behavior is caused by the increase
in the MDC width up to the universal resistivity in the underdoped
region.

\subsection*{Acknowledgment}
We are grateful to Y. Yanase for enlightening discussions. This
work was supported by a Grant-in-Aid for Scientific Research in
Priority Area ``Invention of Anomalous Quantum Materials", and a
Grant-in-Aid for Young Scientists from the Ministry of Education,
Science, Culture, Sports and Technology. Y. A. was supported by
KAKENHI 19674002 and 20030004. ALS is operated by the Department
of Energy's (DOE) Office of Basic Energy Science, Division of
Materials Science. SSRL is operated by the DOE Office of Basic
Energy Science Divisions of Chemical Sciences and Material
Sciences.

\bibliography{ZnLSCO}

\begin{thebibliography}{10}

\bibitem{Fukuzumi}
Y. Fukuzumi, K. Mizuhashi, K. Takenaka, and S. Uchida, Phys. Rev. Lett. {\bf
  76},  684  (1996).

\bibitem{Nachumi}
B. Nachumi, A. Keren, K. Kojima, M. Larkin, G.~M. Luke, J. Merrin, O.
  Tchernyshov, and Y.~J. Uemura, Phys. Rev. Lett. {\bf 77},  5421  (1996).

\bibitem{PanSTM}
S.~H. Pan, E.~W. Hudson, K.~M. Lang, H. Eisaki, S. Uchida, and J.~C. Davis,
  Nature {\bf 403},  746  (2000).

\bibitem{Julien}
M.-H. Julien, T. Feh\'{e}r, M. Horvati\'{c}, C. Berthier, O.~N. Bakharev, P.
  S\'{e}gransan, G. Collin, and J.-F. Marucco, Phys. Rev. Lett. {\bf 84},  3422
   (2000).

\bibitem{Terashima}
K. Terashima, H. Matsui, D. Hashimoto, T. Sato, T. Takahashi, H. Ding, T.
  Yamamoto, and K. Kadowaki, Nature Physics {\bf 2},  27  (2006).

\bibitem{Zabolotnyy}
V.~B. Zabolotnyy, S.~V. Borisenko, A.~A. Kordyuk, J. Fink, J. Geck, A.
  Koitzsch, M. Knupfer, B. Buchner, H. Berger, A. Erb, C.~T. Lin, B. Keimer,
  and R. Follath, Phys. Rev. Lett. {\bf 96},  037003  (2006).

\bibitem{TerashimaZnLSCO}
K. Terashima, T. Sato, K. Nakayama, T. Arakane, T. Takahashi, M. Kofu, and K.
  Hirota, Phys. Rev. B {\bf 77},  092501  (2008).

\bibitem{Nishina}
S. Nishina, T. Sato, T. Takahashi, S.-C. Wang, H.-B. Yang, H. Ding, and K.
  Kadowaki, J. Phys. Chem. Sol. {\bf 63},  1069  (2002).

\bibitem{yoshidaOD}
T. Yoshida, X.~J. Zhou, M. Nakamura, S.~A. Kellar, P.~V. Bogdanov, E.~D. Lu, A.
  Lanzara, Z. Hussain, A. Ino, T. Mizokawa, A. Fujimori, H. Eisaki, C. Kim,
  Z.-X. Shen, T. Kakeshita, and S. Uchida, Phys. Rev. B {\bf 63},  220501(R)
  (2001).

\bibitem{shadow}
Shadow bands seen in the first Brillouin zone in panels (c) and (d) are not
  considered in the present analysis because these bands may be extrinsic
  structures. The origin of the ``shadow band" is disscussed in A. Koitzsch
  {\it et al.}, Phys. Rev. B {\bf 69}, 220505(R) (2004), in terms of structual
  effects.

\bibitem{Dumm}
M. Dumm, S. Komiya, Y. Ando, and D.~N. Basov, Phys. Rev. Lett. {\bf 91},
  077004  (2003).

\bibitem{ZhouNature}
X.~J. Zhou, T. Yoshida, A. Lanzara, P.~V. Bogdanov, S.~A. Kellar, K.~M. Shen,
  W.~L. Yang, F. Ronning, T. Sasagawa, T. Kakeshita, T. Noda, H. Eisaki, S.
  Uchida, C.T. Lin, F. Zhou, J.~W. Xiong, W.~X. Ti, Z.~X. Zhao, A. Fujimori, Z.
  Hussain, and Z.~X. Shen, Nature {\bf 423},  398  (2003).

\bibitem{Mizuhashi}
K. Mizuhashi, K. Takenaka, Y. Fukuzumi, and S. Uchida, Phys. Rev. B {\bf 52},
  R3884  (1995).

\bibitem{AndoHall}
Y. Ando, Y. Kurita, S. Komiya, S. Ono, and K. Segawa, Phys. Rev. Lett. {\bf
  92},  197001  (2004).

\bibitem{OnoHall}
S. Ono, S. Komiya, and Y. Ando, Phys. Rev. B {\bf 75},  024515  (2007).

\bibitem{mdc}
If the surface qualities and the experimental conditions are nearly the same
  between the Zn-free and the Zn-doped samples, the increase of the MDC width
  can be quantitatively attributed to by the Zn impurity scattering.

\bibitem{Mahan}
G.~D. Mahan, {\em Many-Particle Physics}, 2nd ed. (Plenum, New York and London,
  1990).

\bibitem{Kontani}
H. Kontani, K. Kanki, and K. Ueda, Phys. Rev. B {\bf 59},  14723  (1999).

\bibitem{Emery}
V.~J. Emery and S.~A. Kivelson, prl {\bf 74},  3253  (1995).

\bibitem{KMShenPRB}
K.~M. Shen, T. Yoshida, D.~H. Lu, F. Ronning, N.~P. Armitage, W.~S. Lee, X.~J.
  Zhou, A. Damascelli, D.~L. Feng, N.~J.~C. Ingle, H. Eisaki, Y. Kohsaka, H.
  Takagi, T. Kakeshita, S. Uchida, P.~K. Mang, M. Greven, Y. Onose, Y. Taguchi,
  Y. Tokura, Seiki Komiya, Yoichi Ando, M. Azuma, M. Takano, A. Fujimori, and
  Z.-X. Shen, Phys. Rev. B {\bf 69},  054503  (2004).

\bibitem{Okada}
Y. Okada, T. Takeuchi, T. Baba, S. Shin, and H. Ikuta, J. Phys. Soc. Jpn. {\bf
  77},  074714  (2008).

\bibitem{Maki}
H. Won and K. Maki, Phys. Rev. B {\bf 49},  1397  (1994).

\bibitem{TerashimaLSCO}
K. Terashima, H. Matsui, T. Sato, T. Takahashi, M. Kofu, and K. Hirota, Phys.
  Rev. Lett. {\bf 99},  017003  (2007).

\end{thebibliography}

\end{document}